# Spin – polarized quasiparticle injection effects in YBCO thin films

S. Soltan[a, b], J. Albrecht[a, c], and H. -U. Habermeier[*,a]

[a] Max-Planck-Institut für Festkörperforschung, Heisenbergstr. 1, D 70569 Stuttgart, Germany
[b] 1. Physikalisches Institut, Universität Stuttgart, Pfaffenwaldring 57, D-70550 Stuttgart, Germany
[c] Max-Planck-Institut für Metallforschung, Heisenbergstr. 1, D 70569 Stuttgart, Germany

**Abstract**

We report detailed transport studies on ferromagnet-superconductor heterostructures. Epitaxial heterostructures of half-metal colossal magnetoresistive $La_{2/3}Ca_{1/3}MnO_3$ (HM-CMR) and high $T_c$ superconducting $YBa_2Cu_3O_7$ (YBCO) are grown on $SrTiO_3$ (100) single crystal substrates by pulsed laser deposition. Using the HM--CMR layer as source for spin-polarized quasiparticles, we show the effect of injection of spin-polarized quasiparticles into the *ab*-plane and along the *c*-axis of YBCO. The results show a drop in the *ab*-plane resistance $R_{ab}$ (T) in the case of injection along the *c*-axis that is discussed to be related to the opening of a pseudogap.

## 1. Introduction

One of the central issues of debates concerning the normal state of high-$T_c$ superconductors (HTSC) is the physical origin of the so called *pseudogap* at $T_c < T < T^*$; where $T^*$ can be much larger than the critical temperature $T_c$. $T^*$ is the temperature below which a pseudogap in the density of state appears and drastic changes of the spectral density due to strong short-range antiferromagnetic correlations occur, at the same time the long-range phase coherence is absent. This characteristic temperature $T^*$ increases for decreasing doping concentration and is related to the scattering rate or inverse life time ($\tau^{-1}$) of the quasiparticles [1].

For $T_c < T < T^*$, the non-Fermi liquid (NFL) region, it is generally accepted, that between both temperatures unconventional normal-state properties are found. This leads to the model of spin–charge separation proposed by Anderson [2] and supported by Nagaosa and Lee [3,4]. It is assumed that spins are bound together to form spin-singlets and the energy required to split them apart leads to the formation of a spin-gap [4]. Experimental results suggest that $T^*$ is related to the occurrence of a spin-gap in the high- $T_c$ materials [5,6,7].

On the other hand, it is proposed that superconductors with a low carrier density are characterized by a relatively small phase stiffness and thus, the poor screening implies a significant role of phase-fluctuations [1,8-16]. The phase-fluctuation scenario explains quite naturally the strongly enhanced Nernst signal [17] above the critical transition temperature $T_c$ in the underdoped HTSC materials. It was proposed by Emery and Kivelson [1] that the proximity to the Mott-insulating phase implies a strongly reduced phase stiffness $V_o$; where $V_o \sim \rho_s(0)/m^*$ ; with $\rho_s(0)$ being the superfluid density and m* the effective mass, compared to the usual BCS case. This causes the phase ordering temperature $T^{phase} \sim V_o$ to be much lower than the mean-field pair-binding temperature $T_c$. Several authors interpret their experimental results suggesting that the pseudogap is related to the superconducting state; i.e. Cooper-pairs are formed at $T^*$ without phase coherence [17-25].

In this context, spin-polarized quasiparticle injection into HTSC materials is a good candidate in order to test the scattering effects around the pseudogap temperature $T^*$. The injection of quasiparticles into superconductors creates a local nonequilibrium state. One of the simplest methods to probe the pseudogap temperature is the measurement of the dc-resistance in the ab-plane $R_{ab}(T)$. It has been shown that a reduction appears in $R_{ab}(T)$ around T* for the underdoped YBCO compared to the optimally doped material [26,27].

In the past few years, much attention has been paid to junctions consisting of $La_{2/3}Ca_{1/3}MnO_3$ (LCMO), a material that shows a colossal magnetoresistance (CMR) effect, and of $YBa_2Cu_3O_{7-\delta}$ (YBCO) [28-37]. Experiments with these junctions allow us to obtain information about the spin-dependent properties of high-$T_c$ superconductors. It has been suggested by Si [38] that electrons injected from hole-doped rare-earth manganites are a sensitive probe for possible spin-charge separation in high-$T_c$ superconductors.

The high spin polarization is characteristic for hole-doped rare earth manganites. In case of $La_{2/3}Ca_{1/3}MnO_3$, the spin polarization of the transport electrons is theoretically expected to be 100 % and experimentally measured to be 80% [39]. The fact that the in-plane lattice parameters of $La_{2/3}Ca_{1/3}MnO_3$ and $YBa_2Cu_3O_{7-\delta}$ are very similar allows an epitaxial growth of LCMO/YBCO heterostructures with structurally and compositionally sharp interfaces. These heterostructures represent adequate model systems to investigate spin-polarized quasi-particle injection effects into high-temperature superconductors.

Our experimental results are obtained by investigating epitaxial heterostructures of thin films of ferromagnetic LCMO and high-$T_c$ superconducting YBCO. The resistance versus temperature of the YBCO has been determined for c-axis and ab-plane injection of spin-polarized quasiparticles. In these experiments it is found that the superconducting transition temperature of the YBCO film decreases with increasing injection current density. Furthermore, a reduction of the resistance is found around a characteristic temperature $T_d$. Inserting an insulating layer as $SrTiO_3$ in between the ferromagnetic and high-$T_c$ superconducting layer or replacing the ferromagnetic layer by a paramagnetic one leads to the absence of the minimum in the resistance. This difference between both results is related to the loss of spin-polarization by quasiparticle injection across the insulating $SrTiO_3$ layer.

## 2. Experimental Details

Epitaxial heterostructures of $YBa_2Cu_3O_{7-\delta}$, and $La_{2/3}Ca_{1/3}MnO_3$ are grown by pulsed laser deposition. As substrates 10×10 mm$^2$ $SrTiO_3$ (STO) (100) single crystals are used. The preparation conditions are identical for both of the layers. The substrate is kept at a constant temperature of 780 $^\circ$C, the temperature is adjusted and controlled by a far-infrared pyrometric temperature control. The oxygen pressure is 0.4 mbar in case of the LCMO layer deposition and 0.6 mbar in case of the YBCO layer. Afterwards the heterostructure is in–situ annealed for 30-60 minutes at 530$^0$ C in an oxygen pressure of 1.0 bar. This procedure results in films of high crystalline quality, and sharp film–substrate interfaces [28]. Structural studies are carried out using x-ray diffraction (XRD) at room temperature. Two different designs of heterostructure junctions have been used:

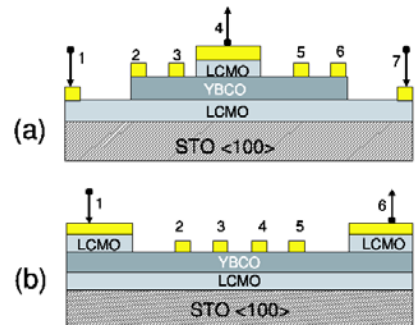

Figure 1: Sketch of the sample geometry chosen for the injection of spin-polarized quasiparticles along the *c*-axis (a) and *ab*-plane (b) of YBCO. [50 nm LCMO, 100 nm YBCO, and 50 nm LCMO] The width of the structure is 1 mm.

(i)*c-axis injection sample design:* The geometry is given in Fig. 1(a). Using conventional lithography we fabricated 3 contact strips for current injection on both LCMO layers with widths of 200μm (contacts numbers 1, 4, and 7), and four contact strips for resistance measurement on the YBCO layer (contacts numbers 2, 3, 5, and 6). As contact materials chromium/gold (Cr/Au) pads with thicknesses of 15 nm and 100 nm are evaporated.

(ii) *ab-plane injection sample design:* The geometry is given in Fig. 1(b). Injection and measurement contacts are made on both, the LCMO top layers, and the free area of the YBCO layer, contact numbers 1 and 6 have been prepared on the LCMO layer. Four contacts are placed on the YBCO layer for resistivity measurements (2,3,4, and 5). As contact materials evaporated chromium/gold (Cr/Au) 15 nm/100 nm are used..

**3. Experimental Results** The x-ray diffraction pattern shows (00ℓ) diffraction peaks for both the LCMO and the YBCO layers, i.e. they are *c*-axis textured [41]. Additionally to XRD, we studied transmission electron microscopy (TEM) images [28]. We can conclude that the bottom LCMO layer and the top YBCO layer grow epitaxially by forming a high-quality interface. With this knowledge we assume that artefacts created due to structural variations at the interface can be neglected.

For spin-polarized quasiparticle injection along the c-axis ( $I_{inj-c}$ ) we use the set-up of Fig. 1a. In the following we inject a current through the YBCO layer $I_{YBCO}$ = 0.1mA, passing through the contacts points 2, 6 and pick up the voltage through 3 and 5. Figure 2a shows the resistance $R_{ab}$ (T) as a function of temperature for different spin-polarized quasiparticle (SPQ) injection currents along the c-axis of YBCO ($I_{inj-c}$ =0.0, 0.3, 0.4, 0.5, 0.75 mA). Starting at room temperature in Fig.2, we see that with increasing injection current density a pronounced enhancement of the resistance at room temperature ($R_{room}$) is found (region I).

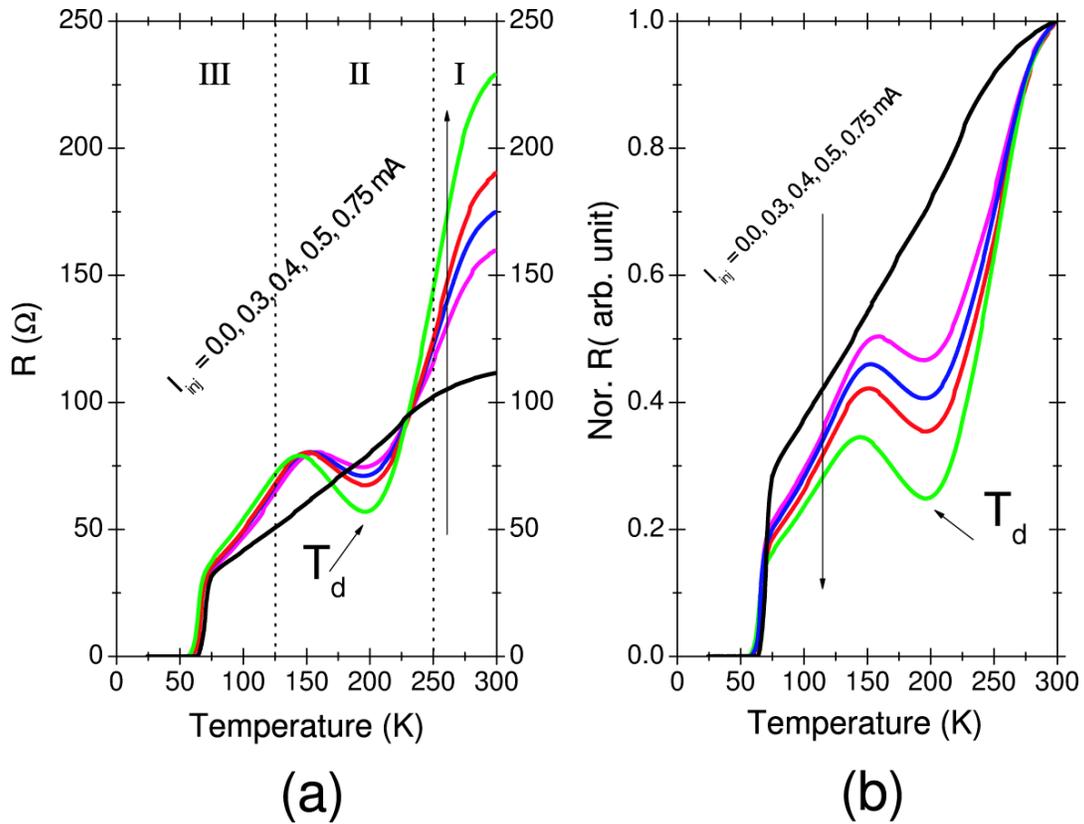

Fig. 2 Temperature-dependent in-plane resistance R (T) for the YBCO layer in Fig (a) for different SPQ currents (a). Note, that an enhancement of the SPQ current causes an increase of the resistance at room temperature, but a drop of the resistance is found around the temperature of the order of 25%. Additionally, a shift from ~63K to ~54K is found in the transition temperature $T_c$. (b) Normalized temperature-dependence of the in-plane resistance.

This enhancement in $R_{room}$ from 112 Ω < $R_{room}$ <225 Ω for 0.0 mA < $I_{inj-c}$ < 0.75 mA, respectively, reflects the scattering process of quasiparticles in the YBCO layer at room temperature. This increase of resistivity is also found in bilayers of LCMO and YBCO without spin injection due to the self-diffusion of spin-polarized quasiparticles [41]. Ee can not completely rule out, however, an additional voltage drop due to the current injection. Nevertheless, this effect is temperature independent therefore we give a rescaled representation normalized to the room temperature resistance in Fig. 2b. Applying the SPQ current, a pronounced drop of the resistance can be seen in region (II). This drop of the resistance from 74 Ω to 55 Ω for $I_{inj}$ = 0.0 mA and $I_{inj}$ = 0.75 mA, respectively, gives rise to a 25% reduction of the resistance around $T_d$, where $T_d$ denotes the temperature at which the minimum of the resistance in the normal state is found (region (II) in Fig. 2a. Region (III) in Fig. 2a, shows the effect of the SPQ on the transition temperature. Here, a shift of the transition temperature $T_c$ to lower values is found; 63.5 K < $T_c$ < 54 K; for an enhancement of the SPQ current density 0.0 mA < $I_{inj-c}$ < 0.75 mA , respectively. This substantial drop of $T_c$ with increasing injection current is regarded as evidence for spin injection into the superconductor[40].

Growing YBCO as top layer onto LCMO leads generally to a decrease of $T_c$ compared to YBCO single layers, also without spin injection. The reduced $T_c$ is considered to have two causes. *First*, the magnetization of the ferromagnet leads to pair-breaking in the superconductor [41]. *Second*, the difference of the chemical potential between both layers leads to an oxygen diffusion at the annealing temperature, this may lead to an underdoped YBCO film.

To obtain more information about the effects of spin-polarized quasiparticle injection another sample geometry is chosen where the current is injected in a different way. We use the configuration of Fig. 1b with SPQ injection into the ab-plane of the YBCO film via 1 and 6, the YBCO current $I_{YBCO}$ = 0.1 mA through 2 and 5 and voltage pick-up through 3 and 4. This configuration allows us to detect the effect of the SPQ into the ab-plane of the YBCO layer. In the important temperature region (II) no effect can be found which is similar to the c-axis injection experiment ( c.f. Fig.2). This is expected to be due to the fact that the SPQ has a finite length $\xi_{FM}$ of spin conservation [41].

More clarification whether the observed effect is related to the SPQ injection into YBCO can be obtained by control samples keeping the thickness and the geometry of the heterostructures constant only changing the source materials for the quasiparticles: *first*, we selected $SrRuO_3$ (SRO) replacing the LCMO layer in the heterostructures. $SrRuO_3$ is a ferromagnet with a lower Curie temperature $T_{Curie}$ = 165 K compared to 1/3-Ca doped LCMO ($T_{Curie}$ = 275 K), and a lower polarization (10-30~%) for the conduction electrons[42]. *Second*, we selected the paramagnetic metal $LaNiO_3$ (LNO) replacing the LCMO layer in the heterostructures. *Third*, we have grown an insulating $SrTiO_3$ (STO) layer in between LCMO and YBCO. The thickness of the STO is ~ 5nm which is sufficient for electronic decoupling. The remarkable difference between all of these experiments is that the drop of the resistance around the temperature $T_d$ is vanishing completely. This drop only appears in case of the injection of *spin-polarized* quasiparticles. We state, that in replacing LCMO by SRO, LNO, or inserting STO layers, *no* reduction of the normal state resistance can be found around the characteristic temperature $T_d$.

## 4. Discussion

In the beginning of the discussion the relevant experimentally obtained results are briefly repeated. 1) The injection of a current through a spin-polarized material into a YBCO film leads to a substantial reduction of the superconducting transition temperature which is less pronounced in case of non-spin polarized contact materials. 2) A distinct reduction of the normal state resistance of the YBCO film occurs around a characteristic temperature $T_d$. 3) This drop in resistance is absent either if the spin-polarized contact material is replaced by a SRO or LNO or if LCMO and YBCO are electronically decoupled by a thin insulating layer.

The shift of $T_c$ with increasing $I_{inj}$ is explained by the following model. Pair breaking due to quasiparticle injection effects (*or proximity effect*) was suggested by Parker [43]: $\Delta(n_{qp})/ \Delta(n_{qp})/\Delta(0) \approx 1 - 2n_{qp}/4N(0) \Delta(0)$; where $\Delta(n_{qp})$ is the energy required to suppress the order parameter of the superconductor due to the density of spin-polarized quasiparticles $n_{qp}$. $N(0)$ and $\Delta(0)$ give the density of states and the order parameter at $T = 0$ K, respectively [44,45]. On the other hand, the proximity effect (i.e. the penetration of Cooper pairs into the neighboring layer) can be ignored for the reduction of $T_c$ in our heterostructures. The large spin exchange energy $J_{spin}$ of the magnetic layer ($J_{spin}$ of LCMO $\approx 3$ eV) [46] prevents Cooper pairs from tunnelling into the LCMO film. Arguments related to the $T_c$ reduction with the pair-breaking model have been elaborated in a previous paper [41].

We concentrate now on the 25 % reduction of R(T) around the temperature $T_d$ due to SPQ injection. In the literature there are two theoretical models that could provide an explanation of the experimental data. We want to discuss the reduction of the resistance in the framework of these two theories:

a) *Phase fluctuations and SPQ injection into HTSC*: We address the point that the SPQ injection could enhance phase fluctuations due to two reasons: *first*: It is known that $T_c^{-1} \sim \tau_{pf}$ where $\tau_{pf}$ is the time scale of Cooper-pair phase fluctuations which might be present below T* ($\tau_{pf} \sim$ ps) [47]. On the other hand, a typical scattering time $\tau_s$ of the SPQ is in the fs range and related to the spin exchange energy $J_{spin}^{-1} \sim \tau_s$ of the SPQ. Due to the short time scale of the SPQ scattering compared to Cooper-pair phase fluctuations the SPQ injection acts as perturbation of the system and thus enhances phase fluctuations. *Second*: If the superconductivity is driven by antiferromagnetic (AFM) correlations [48] and/or precursors of Cooper pairs around T* [1,8,9-16] SPQ injection would also effect these AFM correlations in terms of spin frustration and reorientation. This can lead to an enhancement of Cooper-pair phase fluctuations by SPQ injection.

b) *Spin–charge separation and SPQ injection into HTSC*: Based on the concept of Anderson [2] Si proposed that spin polarized quasiparticles can act as sensitive probes for spin–charge separation [38]. The total resistivity in the normal state of the HTSC material is written as: $\rho_{total} \approx \rho_{charge} + \rho_{spin}$ where $\rho_{spin}$ is the spin resistivity and $\rho_{charge}$ is the charge resistivity [38]. Experimentally, the charge part is measured. The SPQ injection enhances the polarization of the quasiparticles in the $CuO_2$-plane. The spins are paired into singlets. However, the pairs are not static but are fluctuating (quantum spin liquid). The singlet formation explains the appearance of a spin gap and the reduction of spin entropy. The carriers are holes that appear as vacancies in the background of singlet pairs [4]. We relate our results to the fact that the SPQ injection along the *c*-axis of the YBCO layer interrupts the $\rho_{spin}$ part by filling the spin-gap and/or opening additional conduction channels in the $CuO_2$-plane. Hence, this leads to a reduction of the spin resistivity $\rho_{spin}$ and this reduction appears in the experiment. This fits qualitatively to the drop of the resistance around $T_d$ in Fig.3. Further experiments are planned in near future, to see the development of $T_d$ as function of the doping level of YBCO.

## 5. Summary

We have investigated experimentally the effects of spin-polarized quasiparticle (SPQ) injection in heterostructures of manganites and cuprates. From X-ray and HR-TEM measurements we know that $YBa_2Cu_3O_{7-\delta}$ thin films can be grown epitaxially on thin epitaxial films of $La_{2/3}Ca_{1/3}MnO_3$. The transition temperature of the superconducting film decreases with increasing spin-polarized quasiparticles injection density. The SPQ injection along the *c*-axis of the YBCO layer shows a pronounced reduction in the *ab*-plane resistance of the YBCO around a characteristic temperature $T_d$ which is suggested to be the pseudogap temperature. This finding can be explained by spin–charge separation in the normal state of the HTSC materials.

## 6. Acknowledgments

The authors are grateful to G. Cristiani, B. Lemke, and F. Schartner for the preparation of the outstanding samples and designing the lithography masks. We have to thank A. Muramatsu for the suggestion of the in-plane injection sample. (S.S.) is grateful to M. Dressel, A. Muramatsu, M. Dumm, P. Horsch, D. Manske, and M. Mayr for valuable discussion. Also (S.S.) is grateful to Max–Planck–Society and Ministry of Higher Education and Scientific Research, Egyptian government for support.

* Corresponding author: H.-U. Habermeier; email: huh@fkf.mpg.de